\documentclass[11pt]{article} % use larger type; default would be 10pt

\usepackage{amsmath}
\usepackage{amsthm}
\usepackage{amsfonts}
\usepackage{amssymb}
\usepackage[utf8]{inputenc}
\usepackage{geometry} 
\usepackage{graphicx} 
\usepackage{algorithm}
\usepackage{algorithmic}
\usepackage[T1]{fontenc}
\usepackage[utf8]{inputenc}
\usepackage{authblk}

\newtheorem{theorem}{Theorem}[section]

\newtheorem{proposition}[theorem]{Proposition}

\newtheorem{lemma}[theorem]{Lemma}

\theoremstyle{definition}
\newtheorem{definition}[theorem]{Definition}

\newtheorem{remark}[theorem]{Remark}

\newcommand{\C}{\mathcal{C}}
\newcommand{\Uvs}{\mathcal{U}}

\newcommand{\GL}{\textnormal{GL}}
\newcommand{\F}{\mathbb{F}}

\newcommand{\rk}{\textnormal{rk}}
\newcommand{\bs}{\boldsymbol}

\newcommand{\rs}{\mathrm{rs}}

\newcommand{\rowsp}{\textnormal{rs}}

\newcommand{\xx}{x_1,\ldots,x_{k(n-k)}}

\title{On the Genericity of Maximum Rank Distance and Gabidulin Codes\footnote{This work was supported by SNF grant no.\
149716.}}

\author[2]{Alessandro Neri}
\author[1]{Anna-Lena Horlemann-Trautmann}
\author[2]{Tovohery Randrianarisoa}
\author[2]{Joachim Rosenthal}
\affil[1]{EPF Lausanne, Switzerland}
\affil[2]{University of Z{u}rich, Switzerland}

\begin{document}
\maketitle

\begin{abstract}
  We consider linear rank-metric codes in $\F_{q^m}^n$. We show that
  the properties of being MRD (maximum rank distance) and
  non-Gabidulin are generic over the algebraic closure of the
  underlying field, which implies that over a large extension field a
  randomly chosen generator matrix generates an MRD and a
  non-Gabidulin code with high probability. Moreover, we give upper
  bounds on the respective probabilities in dependence on the
  extension degree $m$.
\end{abstract}

\section{Introduction}

Codes in the rank-metric have been studied for the last four
decades. For linear codes a Singleton-type bound can be derived for
these codes. In analogy to MDS codes in the Hamming metric, we call
rank-metric codes that achieve the Singleton-type bound MRD (maximum
rank distance) codes. Since the works of Delsarte \cite{de78} and
Gabidulin \cite{ga85a} we know that linear MRD codes exist for any set
of parameters.  The codes they describe are called Gabidulin codes.

The question, if there are other general constructions of MRD codes
that are not equivalent to Gabidulin codes, has been of large interest
recently. Some constructions of non-Gabidulin MRD codes can be found
e.g.\ in \cite{co15,cr15,sh15}, where many of the derived codes are
not linear over the underlying field but only linear over some
subfield of it. For some small parameter sets, constructions of linear
non-Gabidulin MRD codes were presented in \cite{ho16}.  On the other
hand, in the same paper it was shown that all MRD codes in
$\F_{2^4}^4$ are Gabidulin codes.  In general, it remains an open
question for which parameters non-Gabidulin MRD codes exist, and if
so, how many such codes there are.

In this paper we show that the properties of being MRD (maximum rank
distance) and non-Gabidulin are generic. This implies that over a
large field extension degree a randomly chosen generator matrix
generates an MRD and a non-Gabidulin code with high
probability. Moreover, we give an upper bound on the respective
probabilities in dependence on the extension degree.

The paper is structured as follows. In Section
\ref{sec:preliminaries} we give some preliminary definitions and
results, first for rank-metric codes and then for the notion of
genericity. Section \ref{sec:topology} contains topological results,
showing that the properties of being MRD and non-Gabidulin are
generic. In Section \ref{sec:prob} we derive some upper bounds on the
probability of these two code properties in dependence on the
extension degree of the underlying finite field. We conclude in
Section \ref{sec:conclusion}.

%%%%%%%%%%%%%%%%%%%%%%%%%%%%%%%%%%%%%%%%%%%%%%%%%%%%%%%%%%%%%%%%
%%%%%%%%%%%%%%%%%%%%%%%%%%%%

\section{Preliminaries}\label{sec:preliminaries}

\subsection{Finite Fields and Their Vector Spaces}

The following definitions and results can be found in any textbook on
finite fields, e.g.\ \cite{li94}.  We denote the finite field of
cardinality $q$ by $\F_q$. It exists if and only if $q$ is a prime
power. Moreover, if it exists, $\F_q$ is unique up to isomorphism. An
extension field of extension degree $m$ is denoted by $\F_{q^m}$. If
$\alpha$ is a root of an irreducible monic polynomial in $\F_q[x]$ of
degree $m$, then
$$ \F_{q^m} \cong \F_q[\alpha].$$
We now recall some basic theory on the trace function over finite
fields.
\begin{definition}
  Let $\F_q$ be a finite field and $\F_{q^m}$ be an extension
  field. For $\alpha \in \F_{q^m}$, the \emph{trace} of $\alpha$ over
  $\F_q$ is defined by
  $$\mathrm{Tr}_{\F_{q^m}/\F_q}(\alpha)  :=  \sum_{i=0}^{m-1}\alpha^{q^i}.$$
\end{definition}

For every integer $0<s<m$ with $\gcd(m,s)=1$, we denote by $\varphi_s$
the map given by
 $$ \begin{array}{rcl}
   \varphi_s:\F_{q^m} &\longrightarrow & \F_{q^m} \\
   \alpha & \longmapsto & \alpha^{q^s}-\alpha.
 \end{array}
$$
The following result relates the trace with the maps $\varphi_s$.
	
 \begin{lemma}\label{lem:trace}
   The trace function satisfies the following properties:
   \begin{enumerate}
   \item $\mathrm{Tr}_{\F_{q^m}/\F_q}(\alpha) \in \F_q$ for all
     $\alpha \in \F_{q^m}$.
   \item $\mathrm{Tr}_{\F_{q^m}/\F_q}$ is a linear surjective
     transformation from $\F_{q^m}$ to $\F_q$, where $\F_{q^m}$ and
     $\F_q$ are considered as $\F_q$-vector spaces.
   \item For every $\alpha \in \F_{q^m}^*$, the map
     $\mathrm{T}_{\alpha}$ defined by
   $$ \beta \longmapsto \mathrm{Tr}_{\F_{q^m}/\F_q}(\alpha\beta)$$
   is a linear surjective transformation from $\F_{q^m}$ to $\F_q$,
   where $\F_{q^m}$ and $\F_q$ are considered as $\F_q$-vector spaces.
 \item $\varphi_s$ is a linear transformation from $\F_{q^m}$ to
   itself, considered as $\F_q$-vector space.
 \item For every $s$ coprime to $m$, $\varphi_s(\alpha)=0$ if and only
   if $\alpha \in \F_q$.
 \item $\ker (\mathrm{Tr}_{\F_{q^m}/\F_q})=\mathrm{Im}(\varphi_s)$ for
   every $s$ coprime to $m$ and has cardinality $q^{m-1}$.
 \end{enumerate}

\end{lemma}

\begin{proof}
  The statements of 1., 2.\ and 3.\ can be found e.g.\ in
  \cite[Theorems 2.23 and 2.24]{li94}.
  \begin{enumerate}
  \item[4.] For $\alpha, \beta \in \F_{q^m}$,
    $\varphi_s(\alpha+\beta)=(\alpha+\beta)^{q^s}-(\alpha+\beta)=
    \alpha^{q^s}-\alpha
    +\beta^{q^s}-\beta=\varphi_s(\alpha)+\varphi_s(\beta)$. Moreover,
    for every $\alpha \in\F_{q^m}$, $c\in \F_q$,
    $\varphi_s(\alpha)=c^{q^s}\alpha^{q^s}-c\alpha=c\left(\alpha^{q^s}-\alpha\right)=c\varphi_s(\alpha)$.
  \item[5.] We have $\varphi_s(\alpha)=\alpha^{q^s}-\alpha=0$ if and
    only if $\alpha \in\F_{q^s}$. Since $\alpha \in\F_{q^m}$, this is
    true if and only if $\alpha \in \F_{q^m}\cap \F_{q^s}=\F_q$.
  \item[6.] First we show that $\mathrm{Im}(\varphi_s)\subseteq\ker
    (\mathrm{Tr}_{\F_{q^m}/\F_q})$. Consider an element
    $\alpha\in\mathrm{Im}(\varphi_s)$. Then there exists $\beta\in
    \F_{q^m}$ such that $\alpha=\beta^{q^s}-\beta$. Now
  $$\mathrm{Tr}_{\F_{q^m}/\F_q}(\alpha)=\mathrm{Tr}_{\F_{q^m}/\F_q}(\beta^{q^s}-\beta)=
  \sum_{i=0}^{m-1}(\beta^{q^s}-\beta)^{q^i}=\sum_{i=0}^{m-1}\beta^{q^{s+i}}-\sum_{i=0}^{m-1}\beta^{q^i}.$$
  We observe now that if $i\equiv j \mod m$, then
  $\beta^{q^i}=\beta^{q^j}$. Hence the sum
  $\sum_{i=0}^{m-1}\beta^{q^{s+i}}$ is a rearrangement of
  $\sum_{i=0}^{m-1}\beta^{q^i}$ and
  $\mathrm{Tr}_{\F_{q^m}/\F_q}(\alpha)=0$.  At this point observe that
  the trace function is a polynomial of degree $q^{m-1}$ and so it has
  at most $q^{m-1}$ roots.  This means that $|\ker
  (\mathrm{Tr}_{\F_{q^m}/\F_q})|\leq q^{m-1}$. By part $4$ and $5$ of
  this Lemma
 $$|\mathrm{Im}(\varphi_s)|=\frac{|\F_{q^m}|}{|\ker(\varphi_s)|}=q^{m-1} $$
 and therefore $\mathrm{Im}(\varphi_s)$ and $\ker
 (\mathrm{Tr}_{\F_{q^m}/\F_q})$ must be equal.
\end{enumerate}

\end{proof}

We denote by $\GL_n(q):=\{A\in \F_q^{n\times n} \mid \rk (A) =n\}$ the
general linear group of degree $n$ over $\F_q$.
Furthermore we need the Gaussian binomial $ \binom{n}{k}_q$, which is
defined as the number of $k$-dimensional vector spaces of $\F_q^n$. It
is well-known that
$$  \binom{n}{k}_q = \prod_{i=0}^{k-1} \frac{q^n-q^i}{q^k-q^i}=\frac{\prod_{i=0}^{k-1}(q^n-q^i)}{|\GL_k(q)|}.$$
Moreover, the following fact is well-known and easy to see.
\begin{lemma}\label{lem:intersection}
  Let $k, n$ be two integers such that $0<k\leq n$, and let $\Uvs$ be
  a $k$-dimensional vector subspace of $\F_q^n$.  Then, for every
  $r=0,\ldots,k$, the number of $k$-dimensional subspaces that
  intersect $\Uvs$ in a $(k-r)$-dimensional subspace is
  $$ \binom{k}{k-r}_q  \binom{n-k}{r}_q  q^{r^2} .$$
\end{lemma}
\begin{proof}

  There are $ \binom{k}{k-r}_q$ many subspaces $\Uvs'$ of $\Uvs$ of
  dimension $(k-r)$ that can be the intersection space.
  % For each of these there are $\binom{n-k}{r}_q$ many vector spaces
  % $\Vvs$ in $\Uvs^\perp$ that we can use to complete $\Uvs'$ to a
  % $k$-dimensional space, intersecting $\Uvs$ only in $\Uvs'$.
  Now, in order to complete $\Uvs'$ to a $k$-dimensional vector space,
  intersecting $\Uvs$ only in $\Uvs'$, we have
  $\prod_{i=0}^{r-1}(q^n-q^{k+i})$ choices for the remaining basis
  vectors. For counting how many of these bases span the same space we
  just need to count the number of $k\times k$ matrices of the form
 $$ \left[\begin{array}{cc}
     I_{k-r} & 0 \\
     A  & B
   \end{array}\right],$$
 where $A\in \F_q^{r\times (k-r)}$ and $B\in \GL_r(q)$.  Hence the
 final count is given by
 \begin{align*}
   \binom{k}{k-r}_q\frac{\prod_{i=0}^{r-1}(q^n-q^{k+i})}{q^{r(k-r)}|\GL_r(q)|}&=
   \binom{k}{k-r}_q\frac{q^{kr}\prod_{i=0}^{r-1}(q^{n-k}-q^i)}{q^{r(k-r)}|\GL_r(q)|}\\
   &=\binom{k}{k-r}_q\binom{n-k}{r}_q q^{r^2}.
 \end{align*}
\end{proof}

%%%%%%%%%%%%%%%%%%%%%%%%%%%%%%%%%%%%%%%%%%%%%%%%%

\subsection{Rank-metric Codes}

Recall that there always exists $\alpha\in \F_{q^m}$, such that
$\F_{q^m}\cong \F_q[\alpha] $. Moreover, $\F_{q^m}$ is isomorphic (as
a vector space over $\F_q$) to the vector space $\F_q^m$.
% If not noted differently we will use the isomorphism
% \begin{align*}
%   \F_q^m &\longrightarrow \F_{q^m}\cong \F_q[\alpha] \\
%   (v_1, \dots, v_m) &\longmapsto \sum_{i=1}^m v_i \alpha^{i-1} .
% \end{align*}
One then easily obtains the isomorphic description of matrices over
the base field $\F_q$ as vectors over the extension field, i.e.\
$\F_q^{m\times n}\cong \F_{q^m}^n$.
% Since we will work with matrices over different underlying fields,
% we denote the rank of a matrix $X$ over $\F_q$ by $\rk_q(X)$.

\begin{definition}
  The \emph{rank distance} $d_R$ on $\F_q^{m\times n}$ is defined by
  \[d_R(X,Y):= \rk(X-Y) , \quad X,Y \in \F_q^{m\times n}. \]
  Analogously, we define the rank distance between two elements
  $\boldsymbol x,\boldsymbol y \in \F_{q^m}^n$ as the rank of the
  difference of the respective matrix representations in
  $\F_q^{m\times n}$.
\end{definition}

In this paper we will focus on $ \F_{q^m}$-linear rank-metric codes in
$\F_{q^m}^n$, i.e.\ those codes that form a vector space over $
\F_{q^m}$.
\begin{definition}
  An $\F_{q^m}$-\emph{linear rank-metric code $\mathcal C$} of length
  $n$ and dimension $k$ is a $k$-dimensional subspace of $\F_{q^m}^n$
  equipped with the rank distance.  A matrix $G\in\F_{q^m}^{k\times n}
  $ is called a \emph{generator matrix} for the code $\mathcal C$ if
 $$\mathcal C=\rs(G),$$ where $\rs(G)$ is the subspace generated by the rows of the matrix $G$, called the \emph{row space} of $G$.
\end{definition}

Whenever we talk about linear codes in this work, we will mean
linearity over the extension field $ \F_{q^m}$.  The well-known
Singleton bound for codes in the Hamming metric implies also an upper
bound for codes in the rank-metric:
% We have the following upper bound on the dimension of such codes.
\begin{theorem}\cite[Section~2]{ga85a}
  Let $\mathcal{C}\subseteq \F_{q^m}^{n}$ be a linear matrix code with
  minimum rank distance $d$ of dimension $k$. Then
$$ k\leq n-d+1  .$$
\end{theorem}

\begin{definition}
  A code attaining the Singleton bound is called a \emph{maximum rank
    distance (MRD) code}.
\end{definition}

\begin{lemma}\cite[Lemma 5.3]{ho16}\label{lem:systematic}
  Any linear MRD code $\C \subseteq \F_{q^m}^n$ of dimension $k$ has a
  generator matrix $G \in \F_{q^m}^{k\times n}$ in systematic form,
  i.e.
  $$G = \left[\begin{array}{c|c}
      I_k & X
    \end{array}
  \right]$$ Moreover, all entries in $X$ are from $\F_{q^m} \backslash
  \F_q$.
\end{lemma}

For some vector $(v_1,\dots, v_n) \in \F_{q^m}^n$ we denote the $k
\times n$ \emph{$s$-Moore matrix} by
\[M_{s,k}(v_1,\dots, v_n) := \left( \begin{array}{cccc} v_1 & v_2
    &\dots &v_n \\ v_1^{[s]} & v_2^{[s]} &\dots &v_n^{[s]} \\
    \vdots&&&\vdots \\ v_1^{[s(k-1)]} & v_2^{[s(k-1)]} &\dots
    &v_n^{[s(k-1)]} \end{array}\right) ,\] where $[i]:= q^i$.
\begin{definition}\label{def:Gab}
  Let $g_1,\dots, g_n \in \F_{q^m}$ be linearly independent over
  $\F_q$ and let $s$ be coprime to $m$. We define a \emph{generalized
    Gabidulin code} $\mathcal{C}\subseteq \F_{q^m}^{n}$ of dimension
  $k$ as the linear block code with generator matrix
  $M_{s,k}(g_1,\dots, g_n)$.  Using the isomorphic matrix
  representation we can interpret $\mathcal{C}$ as a matrix code in
  $\F_q^{m\times n}$.
\end{definition}

Note that for $s=1$ the previous definition coincides with the
classical Gabidulin code construction. The following theorem was shown
for $s=1$ in \cite[Section 4]{ga85a}, and for general $s$ in
\cite{ks05}.

\begin{theorem}\label{thm:GabisMRD}
  A generalized Gabidulin code $\mathcal{C}\subseteq \F_{q^m}^{n}$ of
  dimension $k$ over $\F_{q^m}$ has minimum rank
  distance %(over $\F_q$)
  $n-k+1$. Thus generalized Gabidulin codes are MRD codes.
\end{theorem}

The dual code of a code $\mathcal{C}\subseteq \F_{q^{m}}^{n}$ is
defined in the usual way as
\[\mathcal{C}^{\perp} := \{\boldsymbol{u} \in \F_{q^{m}}^{n} \mid
\boldsymbol{u}\boldsymbol{c}^T=0 \quad \forall \boldsymbol{c}\in
\mathcal{C}\}.  \]
% It is well-known that if $\dim(\mathcal{C})=k$, then
% $\dim(\mathcal{C}^{\perp})=n-k$.

In his seminal paper Gabidulin showed the following two results on
dual codes of MRD and Gabidulin codes. The result was generalized to
$s>1$ later on by Kshevetskiy and Gabidulin.

\begin{proposition}\cite[Sections~2 and 4]{ga85a}\cite[Subsection
  IV.C]{ks05}\label{prop:dual1}
  \begin{enumerate}
  \item Let $\mathcal{C}\subseteq \F_{q^{m}}^{n}$ be an MRD code of
    dimension $k$. Then the dual code $\mathcal{C}^{\perp}\subseteq
    \F_{q^{m}}^{n}$ is an MRD code of dimension $n-k$.
  \item Let $\mathcal{C}\subseteq \F_{q^{m}}^{n}$ be a generalized
    Gabidulin code of dimension $k$. Then the dual code
    $\mathcal{C}^{\perp}\subseteq \F_{q^{m}}^{n}$ is a generalized
    Gabidulin code of dimension $n-k$.
  \end{enumerate}
\end{proposition}

% Note that the second result in Proposition \ref{prop:dual1} was not
% stated like this in \cite{ga85a}; Gabidulin showed however that the
% parity check matrix $H\in \F_{q^m}^{(n-k)\times n}$ of a Gabidulin
% code is of the form described in Definition \ref{def:Gab}, which
% implies the statement.
For more information on bounds and constructions of rank-metric codes
the interested reader is referred to \cite{ga85a}.

Denote by $\mathrm{Gal}(\F_{q^m}/\F_q)$ the \emph{Galois group} of
$\F_{q^m}$, consisting of the automorphisms of $\F_{q^m}$ that fix the
base field $\F_q$ (i.e., for $\sigma \in \mathrm{Gal}(\F_{q^m}/\F_q)$
and $\alpha \in \F_q$ we have $\sigma(\alpha) = \alpha$). It is
well-known that $\mathrm{Gal}(\F_{q^m}/\F_q)$ is generated by the
\emph{Frobenius map}, which takes an element to its $q$-th
power. Hence the automorphisms are of the form $x\mapsto x^{[i]}$ for
some $0\leq i \leq m$.

Given a matrix (resp.\ a vector) $A\in \F_{q^m}^{k \times n}$, we
denote by $A^{([s])}$ the component-wise Frobenius $A$, i.e., every
entry of the matrix (resp.\ the vector) is raised to its $q^s$-th
power.  Analogously, given some $\mathcal C \subseteq \F_{q^m}^{k
  \times n}$, we define
$$ \mathcal C^{([s])}:=\left\{\mathbf{c}^{([s])}\mid \mathbf{c}\in \mathcal C \right\}.$$

% We will denote the respective inverse map, i.e.\ the $[i]$-th root,
% by $x\mapsto x^{[-i]}$.

% We denote by $\GL_n(q):=\{A\in \F_q^{n\times n} \mid \rk (A) =n\}$
% the general linear group of degree $n$ over $\F_q$.
The (semi-)linear rank isometries on $\F_{q^m}^{n}$ are induced by the
isometries on $\F_q^{m\times n}$ and are hence well-known, see e.g.\
\cite{be03,mo14,wa96}:
\begin{lemma}\cite[Proposition~2]{mo14}\label{isometries}
  The semilinear $\F_q$-rank isometries on $\F_{q^m}^{n}$ are of the
  form
  \[(\lambda, A, \sigma) \in \left( \F_{q^m}^* \times \GL_n(q) \right)
  \rtimes \mathrm{Gal}(\F_{q^m}/\F_q) ,\] acting on $ \F_{q^m}^n \ni
  (v_1,\dots,v_n)$ via
  \[(v_1,\dots,v_n) (\lambda, A, \sigma) = (\sigma(\lambda
  v_1),\dots,\sigma(\lambda v_n)) A .\] In particular, if
  $\mathcal{C}\subseteq \F_{q^m}^n$ is a linear code with minimum rank
  distance $d$, then
  \[\mathcal{C}' = \sigma(\lambda \mathcal{C}) A \]
  is a linear code with minimum rank distance $d$.
  % Since we want to study isometry classes of linear codes we will
  % neglect $A$ and $H$.
\end{lemma}

% Throughout the paper we apply the Frobenius map to field elements,
% vectors, matrices and subspaces, where we always mean to take the
% Frobenius element-wise.

One can easily check that $\F_q$-linearly independent elements in
$\F_{q^m}$ remain $\F_q$-linearly independent under the actions of
$\F_{q^m}^*, \GL_n(q)$ and $\mathrm{Gal}(\F_{q^m}/\F_q)$. Moreover,
the $s$-Moore matrix structure is preserved under these actions, which
implies that the class of generalized Gabidulin codes is closed under
the semilinear isometries. Thus a code is semilinearly isometric to a
generalized Gabidulin code if and only if it is itself a generalized
Gabidulin code.

In this work we need the following criteria for both the MRD and the
Gabidulin property.  The following criterion for MRD codes was given
in \cite{ho16}, which in turn is based on a well-known result given in
\cite{ga85a}:

% \begin{proposition}\label{lem3}
%   Let $H\in \F_{q^m}^{(n-k)\times n}$ be a parity check matrix of a
%   rank-metric code $\mathcal{C}\subseteq \F_{q^m}^n$. Then
%   $\mathcal{C}$ is an MRD code if and only if
% $$ \rk_{q^m}(VH^T) =n-k$$
% for all $V\in \F_q^{(n-k)\times n}$ with $\rk_{q}(V)=n-k$.
% \end{proposition}
% 
% This criterion is formulated with respect to the parity check matrix
% of a linear code. We can easily derive a criterion for the generator
% matrix of MRD codes from this:
\begin{proposition}\label{prop:MRDCrit}
  Let $G\in \F_{q^m}^{k\times n}$ be a generator matrix of a
  rank-metric code $\mathcal{C}\subseteq \F_{q^m}^n$. Then
  $\mathcal{C}$ is an MRD code if and only if
$$ \rk(VG^T) =k$$
for all $V\in \F_q^{k\times n}$ with $\rk(V)=k$.
\end{proposition}
% \begin{proof}
%   The generator matrix $G$ of $\mathcal{C}$ is a parity check matrix
%   of the dual code $\mathcal{C}^\perp \subseteq \F_{q^m}^n$ of
%   dimension $n-k$. It follows from Proposition \ref{lem3} that
%   $\mathcal{C}^\perp$ is an MRD code if and only if $
%   \rk_{q^m}(VG^T) =k$ for all $V\in \F_q^{k\times n}$ with
%   $\rk_{q}(V)=k$. Since $\mathcal{C}$ is MRD if and only if
%   $\mathcal{C}^\perp$ is MRD (see Proposition \ref{prop:dual1}), the
%   statement follows.
% \end{proof}

Furthermore, we need the following criterion for the generalized
Gabidulin property:

\begin{theorem}\cite[Theorem 4.8]{ho16}\label{thm:GabCrit}
  Let $\mathcal{C}\subseteq \F_{q^m}^n$ be an MRD code of dimension
  $k$. $\mathcal{C}$ is a generalized Gabidulin code if and only if
  there exists $s$ with $\gcd(s,m)=1$ such that
$$ \dim (\mathcal{C} \cap \mathcal{C}^{([s])}) = k-1 .$$
\end{theorem}

% Throughout the paper $I_k$ denotes the identity matrix of size
% $k$. Furthermore, $\langle v_1, \dots, v_n \rangle_{q}$ denotes the
% $\F_q$-vector space generated by $v_1, \dots, v_n$.

%%%%%%%%%%%%%%%%%%%%%%%%%%%%%%%%%%%%%%%%%%%%%%%%%%%%%%%%%%%%%%%%%%%%%%%%%%%

\subsection{The Zariski Topology over Finite Fields}

Consider the polynomial ring $\F_q[x_1,\dots,x_r]$ over the base field
$\F_q$ and denote by $\bar\F_q$ the algebraic closure of $\F_q$, necessarily
an infinite field. For a subset  $S\subseteq \F_q[x_1,\dots,x_r]$ one defines
the algebraic set 
$$   
 V(S): = \{\bs x \in \bar\F_q^r \mid f(\bs x) = 0, \forall f \in S\} .
$$

It is well-known that the algebraic sets inside $\bar\F_q^r$ form the \emph{closed 
sets} of a topology, called the  \emph{Zariski topology}. 
The complements of the Zariski-closed sets are the \emph{Zariski-open} sets.

\begin{definition}
One says that a subset  $G\subset\bar\F_q^r$ defines a \emph{generic set}
if $G$ contains a non-empty Zariski-open set. 
\end{definition}

If the base field are the real number ($\mathbb{R}$)  or complex numbers  ($\mathbb{C}$), 
then a generic set inside $\mathbb{R}^r$ (respectively inside $\mathbb{C}^r$) is necessarily
dense and its complement is contained in an algebraic set of dimension at most $r-1$.

Over a finite field $\F_q$ one has to be a little bit more careful. 
Indeed for every subset $T\subset\F_q^r$ one finds a set of polynomials 
 $S\subseteq \F_q[x_1,\dots,x_r]$ such that 
$$   
\{\bs x \in \F_q^r \mid f(\bs x) = 0, \forall f \in S\} =T.
$$
This follows simply from the fact that a single point inside  $\F_q^r$ forms
a Zariski-closed set and any subset  $T\subset\F_q^r$ is a finite union of points.
However if one has an algebraic set $V(S)$, as defined in the beginning 
of this subsection, then the $\F_{q^m}$-rational points defined through
$$   
 V(S;\F_{q^m}): = \{\bs x \in  \F_{q^m}^r \mid f(\bs x) = 0, \forall f \in S\} 
$$
become in proportion to the whole vector space $\F_{q^m}^r$ thinner and thinner, as
the extension degree $m$ increases. This is a consequence of the Schwartz-Zippel Lemma
which we will formulate, for our purposes, over a finite field. The lemma
itself will be crucial  for our probability
estimations in Section \ref{sec:prob}.

\begin{lemma}[Schwartz-Zippel]\cite[Corollary 1]{sc80}\label{lem:SZ}
  Let $f\in \F_q[x_1,x_2,\dots,x_r]$ be a non-zero polynomial of total
  degree $d \geq 0$. Let  $\F_{q^n}$ be an extension field and let 
$S\subseteq \F_{q^n}$ be a finite set. Let $v_1,
  v_2, \dots, v_r$ be selected at random independently and uniformly
  from $S$. Then
  $$\Pr\big(f(v_1,v_2,\ldots,v_r)=0\big)\leq\frac{d}{|S|}. $$ 
\end{lemma}

%%%%%%%%%%%%%%%%%%%%%%%%%%%%%%%%%%%%%%%%%%%%%%%%%%%%%%%%%%%%%%%%%%%%%%%%%%%

\section{Topological Results}\label{sec:topology}

The idea of this section is to show that the properties of being MRD
and non-Gabidulin are generic properties.
% over the algebraic closure of the respective base field $\F_q$.

Recall that, by Lemma \ref{lem:systematic}, every linear MRD code in
$\F_{q^m}^n$ of dimension $k$ has a unique representation by its
generator matrix $G\in \F_{q^m}^{k\times n}$ in systematic form
$$ G= [\;I_k \mid X\;].$$
Thus we have a one-to-one correspondence between the set of linear MRD
codes in $\F_{q^m}^n$ and a subset of the set of matrices $
\F_{q^m}^{k\times (n-k)}$.  Therefore we want to investigate how many
matrices $X\in \F_{q^m}^{k\times (n-k)}$ give rise to an MRD or a
generalized Gabidulin code, when plugged into the above form of a
systematic generator matrix.

However, to make sense of the definition of genericity, we need to do
this investigation over the algebraic closure of
$\F_{q^m}$. Unfortunately though, some results in the rank-metric, in
particular the definition of and results related to generalized
Gabidulin codes, do not hold over infinite fields. Therefore we will
actually show that the set of matrices fulfilling the criteria of
Corollary \ref{prop:MRDCrit} (for being MRD) and Theorem
\ref{thm:GabCrit} (for being a generalized Gabidulin code) are generic
sets over the algebraic closure.

% \subsection{MRD Codes}
We first show that the set of generator matrices fulfilling the MRD
criterion of Corollary \ref{prop:MRDCrit} is generic.

\begin{theorem}\label{thm:topMRD}
  Let $1\leq k \leq n-1$. The set
$$S_\mathrm{MRD} := \{X \in \bar \F_{q^m}^{k\times (n-k)} \mid \forall A \in \F_q^{n\times k} \textnormal{ of rank } k: \det([I_k \mid X ] A)\neq 0   \} $$
is a generic subset of $ \bar \F_{q^m}^{k\times (n-k)}$.
% The set of $k$-dimensional linear non-MRD codes in $\bar\F_{q^m}^n$
% is a Zariski-closed subset of the set of all $k$-dimensional linear
% codes in $\bar\F_{q^m}^n$.
\end{theorem}
\begin{proof}
  We need to show that $S_\mathrm{MRD}$ contains a non-empty  
  Zariski-open set. In fact we will show that $S_\mathrm{MRD}$ is a non-empty  
  Zariski-open set. The non-empty-ness follows from the existence of
  Gabidulin codes for every set of parameters. Hence it remains to show
  that it is Zariski-open.

  % We show that the set of $k$-dimensional linear non-MRD codes is
  % Zariski-closed, for any $1\leq k \leq n$. Then the statement
  % follows, since the set of all linear non-MRD codes is a finite
  % union (over all possible $k$) of these closed sets.  Recall from
  % Lemma \ref{lem:systematic} that any $k$-dimensional MRD code has a
  % generator matrix of the form $ [I_k \mid X ]$, with $X\in
  % \bar\F_{q^m}^{k\times (n-k)}$.  We now show that the set of
  % $k$-dimensional linear non-MRD codes is a Zariski-closed subset of
  % the set of codes of the form $\rs [I_k \mid X ]$. Since
%$$\left\{\rs [I_k \mid X ] \mid X\in \bar\F_{q^m}^{k\times (n-k)}\right\} \subset \bar{\mathcal{G}}_{q^m}(k,n) ,$$ 
%this then also implies that the set of $k$-dimensional linear non-MRD
% codes in $\bar\F_{q^m}^n$ is a Zariski-closed subset of all
% $k$-dimensional linear codes in $\bar\F_{q^m}^n$..
%
% We know from Corollary \ref{cor3} that $\C =\rs [I_k \mid X ]
% \subseteq \bar\F_{q^m}^n$ is an MRD code if and only if for any $A
% \in \F_q^{n\times k}$ of rank $k$
%$$\det([I_k \mid X ] A)\neq 0 .$$
  If we denote the entries of $X\in \bar\F_{q^m}^{k(n-k)}$ as the
  variables $x_{1},\dots, x_{k(n-k)}$, then, for a given $A \in
  \F_q^{n\times k}$, we have $\det([I_k \mid X ] A) \in
  \F_{q}[x_1,\dots,x_{k(n-k)}]$.
  % We call the product of all these determinants $f_S$; then the set
  % of non-MRD codes is in one-to-one relation with the algebraic set
%$$ \left\{(v_1,\dots,v_{k(n-k)}) \in \bar\F_{q^m}^{k(n-k)} \mid  f_S(v_1,\dots,v_{k(n-k)})= 0 \right\} .$$
%Hence it is a Zariski-closed subset of $\left\{\rs [I_k \mid X ] \mid
%  X\in \bar\F_{q^m}^{k\times (n-k)}\right\} $, which implies the
% statement.
  Hence we can write
  \begin{align*}
   S_\mathrm{MRD}                   & = \mathop{\bigcap_{A \in \F_q^{n\times k}}}_{\rk (A)=k}\{X \in \bar \F_{q^m}^{k\times (n-k)} \mid \det([I_k \mid X ] A)\neq 0   \} \\
                  & = \mathop{\bigcap_{A \in \F_q^{n\times k}}}_{\rk (A)=k} V(\det([I_k \mid X ] A))^C ,
 %                 & = \left(\mathop{\bigcup_{A \in \F_q^{n\times k}}}_{\rk (A)=k} V(\det([I_k \mid X ] A))\right)^C \\
%                  & = V(\{\det([I_k \mid X ] A) \mid A \in \F_q^{n\times k},{\rk (A)=k}\})^C,
%                  & = \mathop{\bigcap_{A \in \F_q^{n\times k}}}_{\rk (A)=k} V(\{\det([I_k \mid X ] A)\})^c,
  \end{align*}
  i.e., it is a finite intersection of Zariski-open sets. Therefore $S_{MRD}$ is a Zariski-open set.
% 
%   It follows from Lemma \ref{lem:open} that $S_\mathrm{MRD}$ is an
%   open set in the Zariski topology.
\end{proof}

% Since the set of MRD codes in $\bar\F_{q^m}^n$ is the complement of
% the set of non-MRD codes, we get:
% \begin{corollary}
%   The set of $k$-dimensional linear MRD codes $\C\subseteq
%   \bar\F_{q^m}^n$ is a Zariski-open, and therefore a generic, subset
%   of the set of all $k$-dimensional linear codes in
%   $\bar\F_{q^m}^n$.
% \end{corollary}
%
% In other words, over the algebraic closure, a randomly chosen
% generator matrix gives rise to an MRD code with high probability.

\begin{remark}
  In Theorem \ref{thm:topMRD} we chose the MRD criterion of Corollary
  \ref{prop:MRDCrit} to show that the MRD property (if seen over some
  finite extension field) is generic. One can do the same by using the
  MRD criterion of Horlemann-Trautmann-Marshall from \cite[Corollary
  3]{ho16}.
\end{remark}

\vspace{0.4cm}
% \subsection{Gabidulin Codes}

We now turn to generalized Gabidulin codes. Firstly we rewrite the
criterion from Theorem \ref{thm:GabCrit} in a more suitable way.

\begin{lemma}\label{lem:reformulation}
  Let $\mathcal{C}\subseteq \F_{q^m}^n$ be an MRD code of dimension
  $k$ and let $0<s<m$ with $\gcd(s,m)=1$. $\mathcal{C}$ is a
  generalized Gabidulin code with parameter $s$ if and only if
  $\rk(X^{(q^s)}-X) = 1$.
\end{lemma}
\begin{proof}
  We know from Theorem \ref{thm:GabCrit} that an MRD code $\C =\rs
  [I_k \mid X] \subseteq \F_{q^m}^n$ is a generalized Gabidulin code
  if and only if $\dim (\mathcal{C} \cap \mathcal{C}^{(q^s)}) =
  k-1$. We get
  \begin{align*}
    \dim (\mathcal{C} \cap \mathcal{C}^{(q^s)}) &= k-1 \\
    \iff \rk \left[ \begin{array}{c|l} I_k & X \\ I_k & X^{(q^s)} \end{array}\right] &= k+1 \\
    \iff \rk \left[ \begin{array}{c|c} I_k & X \\ 0 & X^{(q^s)} - X \end{array}\right] &= k+1\\
    \iff \rk(X^{(q^s)} - X) &= 1 .
  \end{align*}
\end{proof}

The following theorem shows that the set of generator matrices
not fulfilling the generalized Gabidulin criterion of Lemma
\ref{lem:reformulation} is generic over the algebraic closure.

\begin{theorem}\label{thm:topGab}
  Let $1\leq k \leq n-1$ and $0<s<m$ with $\gcd(s,m)=1$. Moreover, let
  $S_\mathrm{MRD}\subseteq \bar \F_{q^m}^{k\times (n-k)}$ be as
  defined in Theorem \ref{thm:topMRD}. The set
$$S_{\mathrm{Gab},s} := \{X \in \bar \F_{q^m}^{k\times (n-k)} \mid \rk(X^{(q^s)}- X) =1  \}\cap S_\mathrm{MRD} $$
is a Zariski-closed subset of the Zariski-open set $S_\mathrm{MRD}$.
% The set of $k$-dimensional generalized Gabidulin codes in
% $\bar\F_{q^m}^n$ is a Zariski-closed subset of the set of all
% $k$-dimensional MRD codes in $\bar\F_{q^m}^n$.
\end{theorem}
\begin{proof}
  % We know from Lemma \ref{lem:reformulation} that an MRD code $\C
  % =\rs [I_k \mid X] \subseteq \bar\F_{q^m}^n$ is a generalized
  % Gabidulin code if and only if $\rk(X^{(q^s)}- X) = 1$ for some
  % $1\leq s\leq m$ coprime to $m$.
  Let $X\in S_{\mathrm{Gab},s}$. Since $X\in S_\mathrm{MRD}$, it
  follows from Lemma \ref{lem:systematic} that $X_{ij}\not\in \F_q$
  for $i=1,\dots, k$ and $j=1,\dots, n-k$.  Then the condition
  $\rk(X^{(q^s)}- X) = 1$ is equivalent to $\rk(X^{(q^s)}- X) < 2$,
  which in turn is equivalent to the condition that all $2\times
  2$-minors of $(X^{(q^s)}-X)$ are zero.
  % i.e.\ if the product of all those determinants is equal to $0$.
  If we denote the entries of $X\in \bar\F_{q^m}^{k(n-k)}$ as the
  variables $x_{1},\dots, x_{k(n-k)}$, then these $2\times 2$-minors
  of $(X^{(q^s)}-X)$ are elements of $
  \F_q[x_1,\dots,x_{k(n-k)}]$. Let us call the set of all these minors
  $S'$.
  % Then the set of generalized Gabidulin codes is in one-to-one
  % relation with the algebraic set
  Then
% $$S_{\mathrm{Gab},s}=  \left\{X \in \bar \F_{q^m}^{k\times (n-k)} \mid  f(x_{1},\dots,x_{k(n-k)})= 0 , \forall f \in S' \right\}\cap S_\mathrm{MRD} .$$
\begin{align*}
 S_{\mathrm{Gab},s} &=  \left\{X \in \bar \F_{q^m}^{k\times (n-k)} \mid  f(x_{1},\dots,x_{k(n-k)})= 0 , \forall f \in S' \right\}\cap S_\mathrm{MRD} \\
                    &=V(S')\cap S_\mathrm{MRD}.
\end{align*}
Hence it is a Zariski-closed subset of $S_\mathrm{MRD}\subseteq \bar
\F_{q^m}^{k\times (n-k)}$.
\end{proof}

Theorem \ref{thm:topGab} implies that the complement of
$S_{\mathrm{Gab},s} $, i.e., the set of matrices that fulfill the MRD
criterion but do not fulfill the generalized Gabidulin criterion, is a
Zariski-open subset of $S_{\mathrm{MRD}} \subset \bar\F_{q^m}^{k\times
  (n-k)}$. Thus, if it is non-empty, then the complement of
$S_{\mathrm{Gab},s} $ is a generic set. The non-empty-ness of this set
will be shown in the following section, in Theorem \ref{thm:main}.
% \begin{corollary}
%   Let $1\leq k \leq n$ and $0<s<m$ with $\gcd(s,m)=1$. Moreover, let
%   $S_\mathrm{MRD}\subseteq \bar \F_{q^m}^{k\times (n-k)}$ be as
%   defined in Theorem \ref{thm:topMRD}. The set
%$$S_{\mathrm{Gab},s} := \{X \in \bar \F_{q^m}^{k\times (n-k)} \mid \rk(X^{(q^s)}- X) =1  \}\cap S_\mathrm{MRD} $$
%is a Zariski-closed subset of $\F_{q^m}^{k\times (n-k)}$.
% \end{corollary}

In other words, over the algebraic closure, a randomly chosen
generator matrix gives rise to a code that does not fulfill the
generalized Gabidulin criterion with high probability.

%%%%%%%%%%%%%%%%%%%%%%%%%%%%%%%%%%%%%%%%%%%%%%%%%%%%%%%%%%%%%%%%%%%%%%%%%%%

\section{Probability Estimations}\label{sec:prob}

In the previous section we have used the Zariski topology to show that
a randomly chosen linear code over $\bar\F_{q^m}$ fulfills most likely
the MRD criterion but not the generalized Gabidulin criterion. Intuitively this
tells us that over a finite, but large, extension field of $\F_{q}$ a
randomly chosen linear code is most likely an MRD code but not a
generalized Gabidulin code. In this section we derive some bounds on
the probability that this statement is true, in dependence of the
field extension degree $m$.

\subsection{Probability for MRD codes}

Here we give a lower bound on the probability that a random linear
rank-metric code in $\F_{q^m}^n$ is MRD. A straight-forward approach
gives the following result.
 
 \begin{theorem}\label{thm:probMRDrough}
   Let $X\in \F_{q^m}^{k(n-k)}$ be randomly chosen. Then
  $$\mathrm{Pr}\big(\; \rs [I_k \mid  X ] \mbox{ is an MRD code } \big)  
  \geq 1-\frac{k\prod_{i=0}^{k-1} (q^n-q^i)}{q^m} \geq 1-kq^{kn-m} .$$
\end{theorem}

 \begin{proof}
   It follows from Corollary \ref{prop:MRDCrit} that $\rs [I_k \mid X
   ] $ is a non-MRD code if and only if
 $$p^*:=\mathop{\prod_{A\in \F_{q}^{n\times k}}}_{\rk(A)=k}  \det([I_k \mid  X ] A) = 0.$$ 
 If we see the entries of $X$ as the variables $x_1,\dots,
 x_{k(n-k)}$, then every variable $x_i$ is contained in at most one
 row of the matrix
  $$[I_k\,|\,X]A = (\sum_{\ell=1}^k A_{ \ell j} + \sum_{\ell=k+1}^n X_{i\ell} A_{\ell j})_{i,j}.$$ 
  Thus $ \det([I_k \mid X ] A) \in \F_q[\xx]$ has degree at most
  $k$. The number of matrices in $ \F_{q}^{n\times k}$ of rank $k$ is
  $\prod_{i=0}^{k-1} (q^n-q^i) \leq q^{kn}$, hence the degree of $p^*$
  is at most $k\prod_{i=0}^{k-1} (q^n-q^i)$.  It follows from Lemma
  \ref{lem:SZ} that
  $$\mathrm{Pr}\big( \;\rs [I_k \mid  X ] \mbox{ is not an MRD code } \big) \leq \frac{\deg p^*}{q^m} $$
  and hence
  $$\mathrm{Pr}\big( \;\rs [I_k \mid  X ] \mbox{ is an MRD code } 
  \big) \geq 1-\frac{\deg p^*}{q^m} \geq 1-\frac{k\prod_{i=0}^{k-1}
    (q^n-q^i)}{q^m} \geq 1-kq^{kn-m} .$$
\end{proof}
 
In the remainder of this subsection we want to improve the bound
obtained in Theorem \ref{thm:probMRDrough}.  To do so we introduce the
set
 $$ \mathcal T(k,n)=\left\{E\in \F_q^{k\times n}\,|\, E \mbox{ is in reduced row echelon form and } \rk(E)=k \right\}.$$
 With this notation we can formulate a variation of Corollary
 \ref{prop:MRDCrit}:
 
 \begin{proposition}\label{prop:improvedMRD}
   Let $G\in \F_{q^m}^{k\times n}$ be a generator matrix of a
   rank-metric code $\mathcal{C}\subseteq \F_{q^m}^n$. Then
   $\mathcal{C}$ is an MRD code if and only if
  $$ \rk(EG^T) =k$$
  for all $E\in \mathcal T(k,n)$.
\end{proposition}

 \begin{proof}
   For every matrix $V\in \F_q^{k\times n}$ consider its reduced row
   echelon form $E_V$. I.e., there exists a matrix $R \in \GL_k(q)$
   such that $V=RE_V$. Then
 $$\det(VG^T)=\det(RE_VG^T)=\det(R)\det(E_VG^T), $$
 and since $\det(R)\neq0$ we obtain that $\rk(VG^T)=k$ if and only if
 $\rk(E_VG^T)=k$. By Corollary \ref{prop:MRDCrit} the statement
 follows.
\end{proof}

For $E\in \mathcal T(k,n)$ we define the polynomial
 $$f_E(x_1,\ldots,x_{k(n-k)}) := \det([I_k\,|\,X]E^T)  \in \F_{q^m}[x_1,\ldots, x_{k(n-k)}] ,$$
 and we furthermore define
  $$ f^*(\xx):= \mathrm{lcm}\left\{f_E(x_1,\ldots,x_{k(n-k)})\,|\, E \in \mathcal T(k,n) \right\},$$
  where, as before, the entries of $X$ are the variables $x_1,\ldots,
  x_{k(n-k)}$.  We can easily observe the following.
 
 \begin{proposition}
   The set of linear non-MRD codes of dimension $k$ in $\F_{q^m}^n$ is
   in one-to-one correspondence with the algebraic set
  $$ V(\{f^*\})=\left\{(v_1,\ldots,v_{k(n-k)}) \in \F_{q^m}^{k(n-k)}\,|\,f^*(v_1,\ldots,v_{k(n-k)})=0\right\} .$$
\end{proposition}
\begin{proof}
  It follows from Proposition \ref{prop:improvedMRD} that the set of
  linear non-MRD codes of dimension $k$ in $\F_{q^m}^n$ is in
  one-to-one correspondence with the algebraic set
  \begin{align*}
    V&=\bigcup_{E\in \mathcal T(k,n)}\left\{(v_1,\ldots,v_{k(n-k)}) \in \F_{q^m}^{k(n-k)}\mid f_E(v_1,\ldots,v_{k(n-k)})=0\right\} \\
    &=\left\{(v_1,\ldots,v_{k(n-k)}) \in \F_{q^m}^{k(n-k)} \mid \prod_{E \in \mathcal T(k,n)} f_E(v_1,\ldots,v_{k(n-k)})=0\right\} \\
    &=\left\{(v_1,\ldots,v_{k(n-k)}) \in \F_{q^m}^{k(n-k)}\mid
      f^*(v_1,\ldots,v_{k(n-k)})=0\right\} ,
  \end{align*}
  where the last two equalities follow from the well-known fact that
$$V(\{f\})\cup V(\{g\})=V(\{fg\})=V(\{\mathrm{lcm}(f,g)\})$$
for any   $ f,g \in \F_q[x_1,\ldots,x_{k(n-k)}]$.
\end{proof}

Note that in the definition of an algebraic set, it suffices to use
the square-free part of the defining polynomial(s). In the above
definition of $V$ however, $f^*(\xx)$ is already square-free, as we
show in the following.
% and let $q(\xx)$ be the square-free of the polynomial $p(\xx)$. Then

 \begin{lemma}
   For every $E\in \mathcal T(k,n)$ the polynomial $f_E(\xx)$ is
   square-free. In particular, the polynomial $f^*(\xx)$ is
   square-free.
 \end{lemma}
 \begin{proof}
   As in the proof of Theorem \ref{thm:probMRDrough}, every variable
   $x_i$ is contained in at most one row of the matrix
   $[I_k\,|\,X]E^T$. % = (\sum_{\ell=1}^k E_{ j \ell} + \sum_{\ell=k+1}^n X_{i\ell} E_{j \ell })_{i,j}.$$
   Hence, in the polynomial $f_E(\xx)$ the degree with respect to
   every variable is at most $1$. Thus $f_E(\xx)$ cannot have multiple
   factors.
 \end{proof}

% 
 % \begin{proposition}
 %   The total degree of the polinomial $p(\xx)$ is at most
%   $$ k\binom{n}{k}_q $$
% \end{proposition}
%  
%  \begin{proof}
%    For every $E \in \mathcal T(k,n)$, every entry of the matrix
%    $[I_k\,|\,X]E$ has total degree at most $1$. Hence $f_E$ has
%    total degree at most $k$. Then
%   $$\deg p\leq \sum_{E\in\mathcal T(k,n)}\deg f_E \leq |\mathcal T(k,n)|k=k\binom{n}{k}_q.$$
% \end{proof}

 We now determine an upper bound on the degree of the defining
 polynomial $f^*$.

 \begin{lemma}\label{lem:deg}
   Let $E\in \mathcal T(k,n)$ and let $\Uvs_0$ be the subspace of
   $\F_q^n$ defined by
  $$ \Uvs_0 := \rs [\; I_k \mid 0 \;]=\left\{(u_1,\ldots,u_n) \in \F_q^n\,|\,u_{k+1}=u_{k+2}=\ldots=u_n=0 \right\}.$$
  Then $$\deg f_E=k-\dim \left(\rowsp(E)\cap \Uvs_0 \right) .$$
\end{lemma}

 \begin{proof}
   Let $r:=k-\dim \left(\rowsp(E)\cap \Uvs_0 \right)$ with $0\leq
   r\leq k$. We can write
  $$E^T=\left[\begin{array}{c}
      E_1 \\
      \hline  
      E_2
    \end{array}
  \right],$$ where $E_1\in \F_q^{k\times k}, E_2\in \F_q^{(n-k)\times
    k}$. Since $\dim \left(\rowsp(E)\cap \Uvs_0 \right)=k-r$, we have
  $\rk(E_2)=r$. Thus there exists a matrix $R\in\GL_k(q)$ such that
  the first $r$ columns of $E_2R$ are linearly independent and the
  last $k-r$ columns are zero. Then
$$f_E(\xx)= \det( [\; I_k \mid X \;] E^T)=\det(R)^{-1}\det(E_1R+XE_2R).$$
The last $k-r$ columns of the matrix $XE_2R$ are zero, i.e., the last
$k-r$ columns of $E_1R+XE_2R$ do not contain any of the variables
$x_i$. On the other hand, the entries of the first $r$ columns are
polynomials in $\F_q[\xx]$ of degree $1$,
since %, since they are linear combinations of $x_i$ that are algebraically independent over $\F_q$.
$$E_1R+XE_2R = \left(\sum_{\ell=1}^n (E_1)_{i\ell} R_{\ell j} + 
  \sum_{\ell=1}^k \sum_{\ell'=1}^n X_{i\ell'} (E_2)_{\ell'\ell}
  R_{\ell j} \right)_{i,j}. $$ Hence we have $\deg f_E\leq r$.
 
% \textbf{In order to show the equality the only proof I thought uses
% the fact that in every row the first $r$ elements are algebraically
% independent and then use induction to show that the elements can not
% cancel. Maybe there is an easier proof.}  Now consider an
% $(n-k)\times (n-k)$ matrix $M$ obtained by completing the first $r$
% columns of $E_2R$ to an invertible matrix. We have that the elements
% in the first $r$ columns of $XE_2R$ and $XM$ coincide. Since $M$ is
% an invertible matrix the entries of $XM$ are algebraically
% independent over $\F_q$ and we get in particular that the elements
% in the first $r$ columns of $XE_2R$ are algebraically independent
% over $\F_q$. At this point consider the set of all $r\times r$
% minors of the first $r$ columns of $XE_2R$. These minors are all
% different and then linearly independent over $\F_q$, otherwise a
% non-trivial linear combination between them would produce a
% non-trivial polynomial relation between the entries of the first
% $XE_2R$.  Now observe that the degree $r$ term of $f_E$ is a linear
% combination of these minors. The coefficents of this linear
% combination are given by the $(k-r)\times (k-r)$ minors of $E_1R$
% where the columns are the last $k-r$, multiplied by $\det(R)^{-1}$.
% Since $E^TR$ has rank $k$ and the last $k-r$ columns of $E_2R$ are
% $0$, it follows that the last $k-r$ columns of $E_1R$ are linear
% independent, and then at least one of the coefficients of the linear
% combination is non-zero. This proves the statement.  \textbf{It's a
% mess, I know. I'm trying to formalize it better.}
Now consider the matrix $E_2R$. We can write

$$E_2R=\left[\begin{array}{c|c}
    \tilde{E}_2 & 0
  \end{array}\right]$$
where $\tilde{E}_2$ is an $(n-k)\times r$ matrix of rank $r$. Hence

$$XE_2R=\left[\begin{array}{c|c}
    X\tilde{E}_2 & 0
  \end{array}\right].$$
First we prove that the entries of the matrix $X\tilde{E}_2$ are
algebraically independent over $\F_q$. Fix $1\leq i \leq k$ and denote
by $(X\tilde{E}_2)_i$ the $i$-th row of the matrix
$X\tilde{E}_2$. Then consider the polynomials $(X\tilde{E}_2)_{ij},$
for $j=1,\ldots, r$, that only involve the variables
$x_{(i-1)(n-k)+1},\ldots, x_{i(n-k)}$ . The Jacobian of these
polynomials is $\tilde{E}_2^T$, whose rows are linearly independent
over $\F_q$. Therefore the elements in every row are algebraically
independent over $\F_q$.  Moreover different rows involve different
variables, hence we can conclude that the entries of the matrix
$X\tilde{E}_2$ are algebraically independent over $\F_q$.

At this point consider the set of all $r\times r$ minors of
$X\tilde{E}_2 $. These minors are all different and hence linearly
independent over $\F_q$, otherwise a non-trivial linear combination of
them that gives $0$ would produce a non-trivial polynomial relation
between the entries of $X\tilde{E}_2R$.  Now observe that the degree
$r$ term of $f_E$ is a linear combination of these minors.  If we
write
$$ E_1R=\left[\begin{array}{c|c}
    * & \tilde{E}_1 
  \end{array}\right],$$
where $\tilde{E}_1\in \F_q^{k\times (k-r)}$, then the coefficients of
this linear combination are given by the $(k-r)\times (k-r)$ minors of
$\tilde{E}_1$, multiplied by $\det(R)^{-1}$.  Since $E^TR$ has rank
$k$ and the last $k-r$ columns of $E_2R$ are $0$, it follows that the
columns of $\tilde{E}_1$ are linearly independent, and hence at least
one of the coefficients of the linear combination is non-zero. This
proves that the degree $r$ term of $f_E$ is non-zero, and hence $\deg
f_E=r$.
\end{proof}

We can now give the main result of this subsection, an upper bound on
the probability that a random generator matrix generates an MRD code:
 
 \begin{theorem}\label{thm:probMRD}
   Let $X\in \F_{q^m}^{k(n-k)}$ be randomly chosen. Then
  $$\mathrm{Pr}\big( \;\rs [I_k \mid  X ] \mbox{ is an MRD code } \big) \geq 1-\sum_{r=0}^kr\binom{k}{k-r}_q\binom{n-k}{r}_qq^{r^2}q^{-m}  .$$
\end{theorem}
 
\begin{proof}
  % The total degree of the polynomial $p(\xx)$ is at most
 % $$\sum_{r=0}^kr\binom{k}{k-r}_q\binom{n-k}{r}_qq^{r^2}$$
 % (\textbf{I'm trying to adapt the $q$-Vandermonde identity to find a
 % nicer formula. Not sure it's possible.})
 % and hence
  For every $r=0,1,\ldots,k$ we define the set
  $$ \mathcal T_r=\left\{E\in \mathcal T(k,n)\,|\,\dim \left(\Uvs_0\cap \rowsp(E)\right)=k-r\right\},$$
  where
  $$ \Uvs_0 := \rs [\; I_k \mid 0 \;]=\left\{(u_1,\ldots,u_n) \in \F_q^n\,|\,u_{k+1}=u_{k+2}=\ldots=u_n=0 \right\}.$$
  By Lemma \ref{lem:intersection} we have
  $$\left|\mathcal T_r \right|=\binom{k}{k-r}_q\binom{n-k}{r}_qq^{r^2}.$$
  Moreover, by Lemma \ref{lem:deg}, if $E\in\mathcal T_r$, then $\deg
  f_E=r$.  Hence, by the definition of $f^*(\xx)$, we have
  $$\deg f^*\leq \sum_{E\in\mathcal T(k,n)}\deg f_E=\sum_{r=0}^k\sum_{E \in \mathcal T_r}\deg f_E=\sum_{r=0}^k r\binom{k}{k-r}_q\binom{n-k}{r}_qq^{r^2}.$$
  With Lemma \ref{lem:SZ}, the statement follows.
  % we get
  %$$\mathrm{Pr}\left[ \rs [I_k \mid  X ] \mbox{ is an MRD code } \right] \geq 1-\frac{\deg f^*}{q^m}, $$
  %which implies the statement.
\end{proof}

Remember that we know how to construct MRD codes, namely as Gabidulin
codes, for any set of parameters. Hence the probability that a
randomly chosen generator matrix generates an MRD code is always
greater than zero. However, the lower bound of Theorem
\ref{thm:probMRD} is not always positive. In particular, for
 $$m<k(n-k)+\log_qk$$
 % $$m<\log_q k(n-k)+1$$
 we get
 % $$1-\sum_{r=0}^kr\binom{k}{k-r}_q\binom{n-k}{r}_q
 % q^{r^2}q^{-m} $$ $$= 1-q^{-m}\left( k(n-k)q +
 %   \sum_{r=2}^kr\binom{k}{k-r}_q\binom{n-k}{r}_q q^{r^2}
 % \right)<0,$$
 \begin{align*}
   &1-\sum_{r=0}^kr\binom{k}{k-r}_q\binom{n-k}{r}_q  q^{r^2}q^{-m} \\
   = &1-q^{-m}\left( k\binom{n-k}{k}_qq^{k^2} + \sum_{r=1}^{k-1}r\binom{k}{k-r}_q\binom{n-k}{r}_q  q^{r^2} \right)\\
   \leq& 1-q^{-m}\left( kq^{k(n-k)}\right)< 0,
 \end{align*}
 i.e., the bound is not tight (and not sensible) in these cases.
 % \begin{remark}
 %   Observe that the quantity
%  $$\sum_{r=0}^kr\binom{k}{k-r}_q\binom{n-k}{r}q^{r^2}$$
%  is a polynomial in $\mathbb N[q]$ of degree $k(n-k)$ with leading
%  coefficient $k$. Hence, a necessary condition for
%  $$1-\sum_{r=0}^kr\binom{k}{k-r}_q\binom{n-k}{r}q^{r^2}q^{-m}>0$$ is that $$m>k(n-k)+\log_qk$$
% \end{remark}

\begin{figure}[ht]
   \begin{center}
     \includegraphics[width=7.25cm]{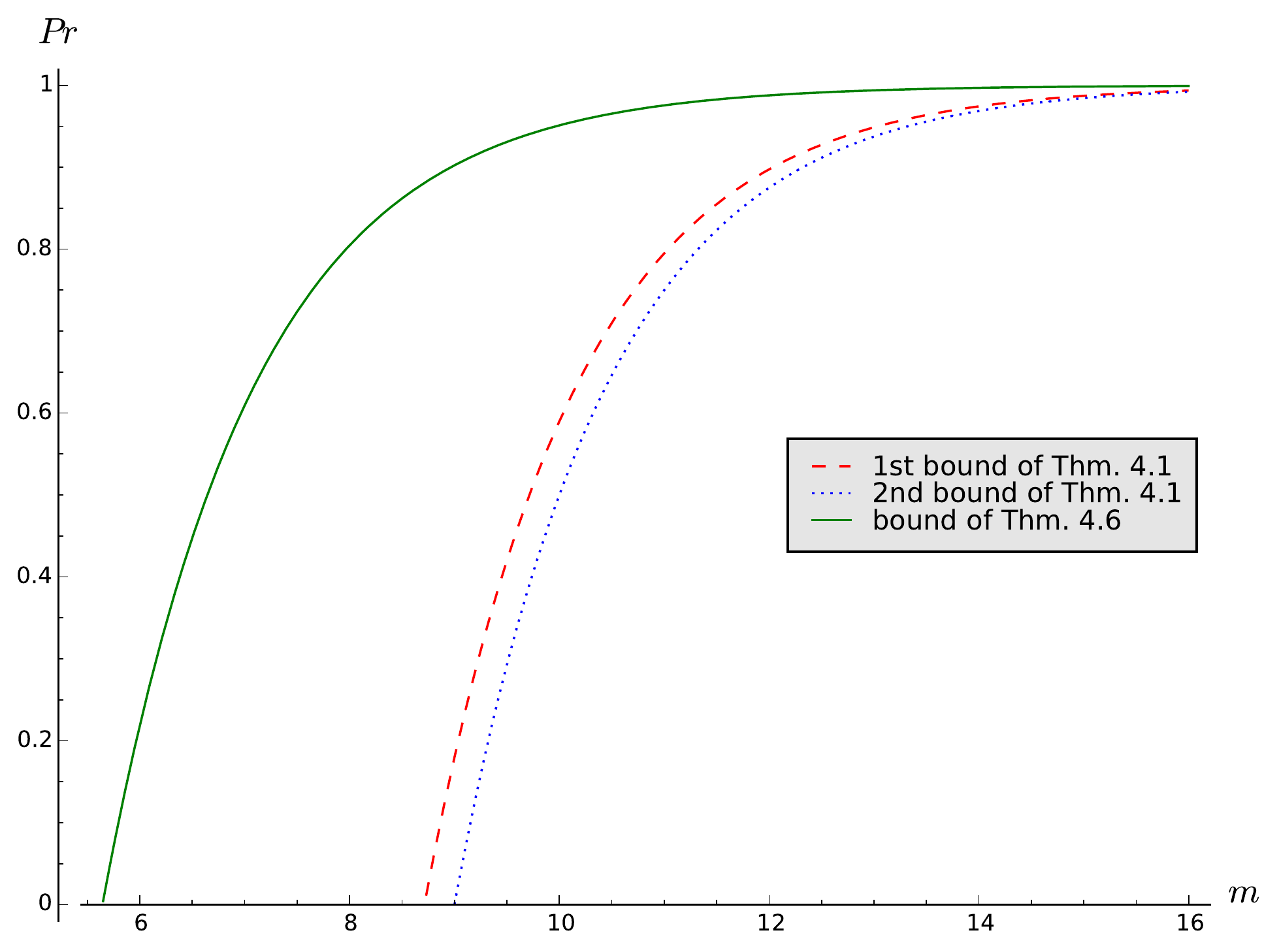}
     % \end{center}
     % \caption{Lower bounds on the probability that a randomly chosen
     % generator matrix in $\F_{2^m}^{2\times 4}$ generates an MRD
     % code.}
     % \label{plotMRD1}
     % \end{figure}
%
     % \begin{figure}
     %   \begin{center}
     \includegraphics[width=7.25cm]{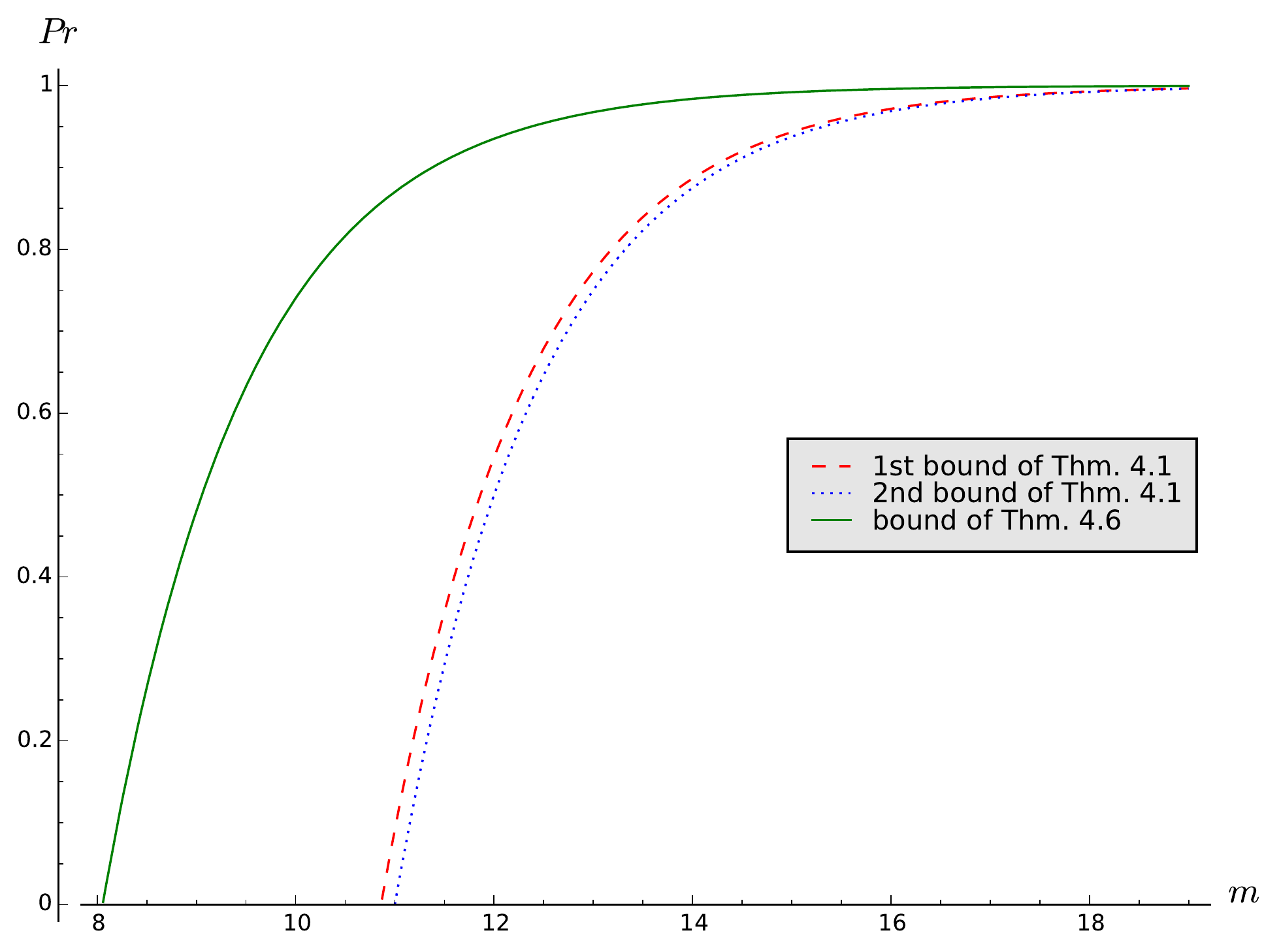}
   \end{center}
   \caption{Lower bounds on the probability that a randomly chosen
     generator matrix in $\F_{2^m}^{2\times 4}$ (left) and
     $\F_{2^m}^{2\times 5}$ (right) generates an MRD code.}
   \label{plotMRD2}
 \end{figure}

 Figure \ref{plotMRD2} depicts the lower bounds of Theorem
 \ref{thm:probMRDrough} and Theorem \ref{thm:probMRD} for small
 parameters. One can see that the bounds of Theorem \ref{thm:probMRD}
 really is an improvement over the bound of Theorem
 \ref{thm:probMRDrough}.

 % \begin{figure}
 %   \begin{center}
 %     \includegraphics[width=10cm]{probs_MRD2_n4k2q3.pdf}
 %   \end{center}
 %   \caption{Lower bounds on the probability that a randomly chosen
 %   generator matrix in $\F_{3^m}^{2\times 4}$ generates an MRD
 %   code.}
 %   \label{plotMRD3}
 % \end{figure}
 
%%%%%%%%%%%%%%%%%%%%%%%%%%%%%%%%%%%%%%%%%%%%%%%%%%%%%%%%%%%%%%%%%%%%%%%%%%%

 \subsection{Probability for Gabidulin codes}

 We have seen in Theorem \ref{thm:topGab} that the set of matrices in
 $\F_{q^m}^{k\times n}$ in systematic form that generate a generalized
 Gabidulin code with parameter $s$ (such that $0<s<m$ with
 $\gcd(s,m)=1$) is in one-to-one correspondence with a subset of the
 set
 $$\left\{X \in \F_{q^m}^{k \times (n-k)}\,| \; \rk(X^{(q^s)}-X)=1 \right\}, $$
 namely with the elements that represent an MRD code. By Lemma
 \ref{lem:systematic} we furthermore know that, if $X$ has entries
 from $\F_q$, then $\rs [\;I_k\mid X\;]$ is not MRD. Hence the set of
 matrices in systematic form that generate a Gabidulin code is in
 one-to-one correspondence with a subset of the set
%  $$\mathcal{G}(1):=\left\{X \in (\F_{q^m}\smallsetminus \F_q)^{k \times (n-k)}\,| \; \rk(X^{(q)}-X)=1 \right\}.$$
%  \left\{G=[I_k\,|\, X]\,: \; G \mbox{ generates a Gabidulin code }
%  \right\}= \left\{G=[I_k\,|\, X]\,: \; G \mbox{ generates an MRD
%    code } \right\}\cap In general, for every integer $0<s<m$ with
%  $\gcd(s,m)=1$, the set of matrices that generate a generalized
%  Gabidulin code with parameter $s$ is a subset of
 $$ \mathcal G(s):=\left\{X \in (\F_{q^m}\smallsetminus \F_q)^{k
     \times (n-k)}\,| \; \rk(X^{(q^s)}-X)=1 \right\}.$$
 For simplicity we make the following estimation of the probability
 that a randomly chosen generator matrix generates a generalized
 Gabidulin code.
 
 \begin{lemma}\label{lem:probGabunion}
   Let $X\in\F_{q^m}^{k \times (n-k)}$ be randomly chosen.
   % Denote by $\phi(m)$ the Euler-$\phi$-function.
   Then
 $$\mathrm{Pr}\big( \;\rs[I_k\,|\,X] \mbox{ is a gen.\ Gabidulin code } \big)\leq 
 \mathop{\sum_{0<s<m}}_{\gcd(s,m)=1}\mathrm{Pr}\big(X\in\mathcal
 G(s)\big) =\mathop{\sum_{0<s<m}}_{\gcd(s,m)=1}\frac{|\mathcal
   G(s)|}{q^{mk(n-k)}} .$$
 % = \phi(m) \frac{|\mathcal G(s)|}{q^{mk(n-k)}}.$$
\end{lemma}

\begin{proof}
  The inequality follows from the fact that the set of generalized
  Gabidulin codes is in one-to-one correspondence with a subset of the
  set
$$\mathop{\bigcup_{0<s<m}}_{\gcd(s,m)=1}  \mathcal G(s) . $$
Since $|\F_{q^m}^{k(n-k)}| = q^{mk(n-k)}$, the statement follows.
% $$ \mathop{\sum_{0<s<m}}_{\gcd(s,m)=1}\mathrm{Pr}[X\in\mathcal G(s)]
% =\mathop{\sum_{0<s<m}}_{\gcd(s,m)=1}\frac{|\mathcal
% G(s)|}{q^{mk(n-k)}}.$$
\end{proof}

For every integer $0<s<m$ with $\gcd(m,s)=1$, we now define the map
$\varPhi_s$ given by
 $$\begin{array}{rcl}
   \varPhi_s:\F_{q^m}^{k\times (n-k)} &\longrightarrow & \F_{q^m}^{k\times (n-k)} \\
   X & \longmapsto & X^{(q^s)}-X.
 \end{array}
$$
Observe that $\varPhi_s$ is exactly the function that maps every entry
$X_{ij}$ of the matrix $X$ to $\varphi_s(X_{ij})$.
Moreover we define the sets
\begin{align*}
  \mathcal R_1 &: =\left\{A\in \F_{q^m}^{k\times (n-k)}\,|\, \rk(A)=1\right\}, \\
  \mathcal R_1^*&:=\left\{A\in (\F_{q^m}^*)^{k\times (n-k)}\,|\, \rk(A)=1\right\}, \\
  \mathcal K
  &:=\left(\ker\left(\mathrm{Tr}_{\F_{q^m}/\F_q}\right)\right)^{k\times(n-k)}
  .
\end{align*}

We state now the crucial results that will help us to compute an upper
bound on the cardinality of the sets $\mathcal G(s)$.

\begin{lemma}\label{lem:phi}
 
  \begin{enumerate}
  \item Given a matrix $A \in \F_{q^m}^{k\times (n-k)}$, there exists
    a matrix $X \in \F_{q^m}^{k\times (n-k)}$ such that
    $\varPhi_s(X)=A$ if and only if $A\in\mathcal K$.
  \item If $A \in \mathcal R_1$, then
  $$|\varPhi_s^{-1}(A)|=\begin{cases}
    0 &  \mbox{ if } A\notin \mathcal K \\
    q^{k(n-k)} & \mbox{ if } A\in \mathcal K.
  \end{cases}
$$
\item For every integer $s$ coprime to $m$ $$\mathcal
  G(s)=\varPhi_s^{-1}(\mathcal R_1^*\cap \mathcal K),$$
  % where $\mathcal K$ is the set of all $k\times (n-k)$ matrices with
  % all the entries in $\ker\left(\mathrm{Tr}_{\F_{q^m}/\F_q}\right)$,
  and
  $$|\mathcal G(1)|=|\mathcal G(s)|=q^{k(n-k)}|\mathcal R_1^*\cap \mathcal K|.$$

%  
  % \item For every integer $s$ coprime to $m$
%  $$|\mathcal G(1)|=|\mathcal G(s)|=q^{k(n-k)}|\mathcal R_1^*\cap \mathcal K|,$$
%  where $\mathcal K$ is the set of all $k\times (n-k)$ matrices with
%  all the entries in $\ker\left(\mathrm{Tr}_{\F_{q^m}/\F_q}\right)$
\end{enumerate}
\end{lemma}

\begin{proof}
  \begin{enumerate}
  \item Since $\varPhi_s$ is the function that maps every entry
    $X_{ij}$ of the matrix $X$ to $\varphi_s(X_{ij})$, we have that $A
    \in \mathrm{Im}(\varPhi_s)$ if and only if every entry $A_{ij}$ of
    $A$ belongs to $\mathrm{Im}(\varphi_s)$. By Lemma \ref{lem:trace}
    part $6$ this is true if and only if every $A_{ij}$ belongs to
    $\ker\left(\mathrm{Tr}_{\F_{q^m}/\F_q}\right)$.
  
  \item If $A\notin \mathcal K$, then by part $1$ this means that
    $\varPhi_s^{-1}(A)=\emptyset$. Otherwise, again by part $1$,
    $\varPhi_s^{-1}(A)\neq\emptyset$. In this case every entry
    $A_{ij}$ belongs to $\mathrm{Im}(\varphi_s)$, and since
    $\varphi_s$ is linear over $\F_q$,
    $|\varphi_s^{-1}(A_{ij})|=|\ker(\varphi_s)|$. Since, by Lemma
    \ref{lem:trace},
    $$ |\ker(\varphi_s)|=\frac{|\F_{q^m}|}{|\mathrm{Im}(\varphi_s)|}=q,$$
    and $A$ has $k(n-k)$ entries, we get
    $|\varPhi_s^{-1}(A)|=q^{k(n-k)}$.
   
  \item
    % Let $\left(\F_{q^m}^*\right)^{k\times (n-k)}$ be the set of all
    % the $k \times (n-k)$ matrices over $\F_{q^m}$ with all non-zero
    % entries.
    Observe that $\mathcal R_1^*=\mathcal R_1\cap
    \left(\F_{q^m}^*\right)^{k\times (n-k)}$. Moreover
   $$ \varPhi_s^{-1}(\mathcal R_1)=\left\{X \in \F_{q^m}^{k \times (n-k)}\,| \; \rk(X^{(q^s)}-X)=1 \right\}$$
   and, by Lemma \ref{lem:trace} part $5$,
   $$\varPhi_s^{-1}(\left(\F_{q^m}^*\right)^{k\times (n-k)})=\left(\F_{q^m}\smallsetminus \F_q\right)^{k \times (n-k)}. $$
   Hence
   $$
   \varPhi_s^{-1}(\mathcal R_1^*)=\varPhi_s^{-1}(\mathcal R_1\cap
   \left(\F_{q^m}^*\right)^{k\times (n-k)})=\varPhi_s^{-1}(\mathcal
   R_1)\cap \varPhi_s^{-1}(\left(\F_{q^m}^*\right)^{k\times
     (n-k)})=\mathcal G(s).
 $$ 
 Now we can write
   $$ \mathcal R_1^*=(\mathcal R_1^*\cap\mathcal K)\cup(\mathcal R_1^*\cap \mathcal K^c)$$
   and by part $1$ we have that $\varPhi_s^{-1}(\mathcal R_1^*\cap
   \mathcal K^c)=\emptyset$. Then
   $$\mathcal G(s)= \varPhi_s^{-1}(\mathcal R_1^*)=
   \varPhi_s^{-1}(\mathcal R_1^*\cap \mathcal
   K)\cup\varPhi_s^{-1}(\mathcal R_1^*\cap \mathcal
   K^c)=\varPhi_s^{-1}(\mathcal R_1^*\cap \mathcal K).$$ By part $2$
   we have $\left|\varPhi_s^{-1}(\mathcal R_1^*\cap \mathcal
     K)\right|=q^{k(n-k)}|\mathcal R_1^*\cap \mathcal K|$, which
   proves the statement.
 \end{enumerate}

\end{proof}
% \begin{corollary}
%   For every integer $0<s<m$ with $(s,m)=1$, $|\mathcal
%   G(1)|=|\mathcal G(s)|$ and hence
%   $$\mathbb P[\{[I_k\,|\,X] \mbox{ generates a gen. Gabidulin code }\}]\leq\phi(m)\mathbb P[\mathcal G(1)].$$
% \end{corollary}
%  

% At this point we need only to estimate the cardinality of the set
% $\mathcal G(1)$.
In analogy to the previous subsection we now first derive a
straight-forward upper bound on the probability that a random
generator matrix gives rise to a generalized Gabidulin
code. Afterwards we will improve this bound.

\begin{theorem}\label{thm:probGabrough}
  Let $X\in \F_{q^m}^{k(n-k)}$ be randomly chosen. Denote by $\phi(m)$
  the Euler-$\phi$-function. Then
  $$\mathrm{Pr}\big( \;\rs [I_k \mid  X ] \mbox{ is a generalized Gabidulin code } \big) 
  \leq \phi(m) ( 2q^{1-m})^{\lfloor\frac{k}{2}\rfloor
    \lfloor\frac{n-k}{2}\rfloor}$$
\end{theorem}
\begin{proof}
  We want to derive the cardinality of $\mathcal G(s)$ for any valid
  $s$.  For this, by Lemma \ref{lem:phi} part $3$, we note that these
  cardinalities are all equal to the cardinality of $\mathcal G(1)$.
  Now for any $X \in (\F_{q^m}\smallsetminus \F_q)^{k \times (n-k)}$
  the rank of $X^{(q)}-X$ is greater than zero. Therefore we can also
  write
$$ \mathcal G(1)=\left\{X \in (\F_{q^m}\smallsetminus \F_q)^{k \times (n-k)}\,| \; \rk(X^{(q)}-X)\leq 1 \right\}.$$
The condition that $\rk(X^{(q)}-X)\leq 1$ is equivalent to that any
$2\times 2$-minor of $X^{(q)}-X$ is zero. Hence a necessary condition
is that any set of non-intersecting minors is zero. We have
$\lfloor\frac{k}{2}\rfloor \lfloor\frac{n-k}{2}\rfloor$ many such
non-intersecting minors, each of which has degree at most $2q$ if we
see the entries of $X$ as the variables $x_1,\dots, x_{k(n-k)}$. With
Lemma \ref{lem:SZ} we get for each minor $M_{ij}$,
$$\Pr(M_{ij} = 0) \leq 2q^{1-m}. $$
% or equivalently
% $$\Pr(M_{ij} \neq 0) \geq 1-2q^{1-m} .$$
Since the non-intersecting minors are independent events, the
probability that all of these are zero is at most
$$( 2q^{1-m})^{\lfloor\frac{k}{2}\rfloor \lfloor\frac{n-k}{2}\rfloor}.$$
With Lemma \ref{lem:probGabunion} and the fact that the number of $s$
with $\gcd(s,m)=1$ is given by $\phi(m)$, the statement follows.
\end{proof}

% However, the bound obtained for the probability that a randomly
% chosen code is a generalized Gabidulin code can be improved.
To improve the above bound we need the following lemma.

 \begin{lemma}\label{lem:R1}
   The set $\mathcal R_1^*\cap \mathcal K$ is in one-to-one
   correspondence with the set
   \begin{align*}
     V_R :=& \left\{\left(\bs\alpha,\bs\beta\right)\in \F_{q^m}^k
       \times \F_{q^m}^{n-k-1}\,|\,\alpha_i, \alpha_i\beta_j \in
       \ker \left(\mathrm{Tr}_{\F_{q^m}/\F_q}\right)\smallsetminus\{0\}\right\} \\
     =& \left\{\left(\bs\alpha,\bs\beta\right)\in \F_{q^m}^k \times
       \F_{q^m}^{n-k-1}\,|\,\alpha_i, \in \ker
       \left(\mathrm{Tr}_{\F_{q^m}/\F_q}\right)\smallsetminus\{0\},
       \beta_j \in \bigcap_{i=1}^k\ker
       \left(\mathrm{T}_{\alpha_i}\right)\smallsetminus\{0\}\right\}
   \end{align*}
   via the map $\psi:V_R \longrightarrow \mathcal R_1^*\cap \mathcal
   K$, given by
  $$ (\bs\alpha,\bs\beta)\longmapsto\left[\begin{array}{c}\alpha_1 \\
      \vdots \\
      \alpha_k    
    \end{array}\right]\left[1, \beta_1,\ldots, \beta_{n-k-1}\right],
$$
and hence
$$|\mathcal R_1^*\cap \mathcal K|\leq (q^{m-1}-1)^{n-1} $$
\end{lemma}
  
\begin{proof}
  From the definition of the set $V_R$ it is clear that the map $\psi$
  is well-defined, i.e., it maps every element in $V_R$ to an element
  in $\mathcal R_1^*\cap \mathcal K$.
 
  Let $(\bs\alpha,\bs\beta)$, $(\bs\gamma,\bs\delta)$ be two elements
  that have the same image.  Then the first column of
  $\psi(\bs\alpha,\bs\beta)$ and the first column of
  $\psi(\bs\gamma,\bs\delta)$ are equal, hence $\bs\alpha=\bs\gamma$.
  Also the first rows of $\psi(\bs\alpha,\bs\beta)$ and
  $\psi(\bs\gamma,\bs\delta)$ are equal, thus
  $\alpha_1\beta_j=\gamma_1\delta_j$ for every $j=1,\ldots, n-k-1$,
  and since $\alpha_1=\gamma_1\neq 0$ we get $\bs\beta=\bs\delta$ and
  this shows the injectivity of the map $\psi$.
 
  In order to show the surjectivity consider a rank $1$ matrix
  $A\in\mathcal R_1^*\cap \mathcal K$ with entries $A_{ij}$.  Consider
  the vectors $\bs\alpha=(A_{11},\ldots,A_{k1})^T$ and
 $$\bs\beta=A_{11}^{-1}(A_{12},\ldots,A_{1(n-k)})^T.$$
 It is clear that $(\bs\alpha, \bs\beta)\in V_R$, and that
 $\psi(\bs\alpha,\bs\beta)=A$.
 
 At this point for every $\alpha_i$ we have $q^{m-1}-1$ possible
 choices, while for every $\beta_i$ we have a number of choices that
 is less or equal to $|\ker(T_{\alpha_1})\smallsetminus\{0\}|$, that
 is again $q^{m-1}-1$. Therefore we get
 $$|\mathcal R_1^*\cap \mathcal K|\leq (q^{m-1}-1)^{n-1}.$$
\end{proof}

We can now formulate the main result concerning the probability that a
random linear rank-metric code is a generalized Gabidulin code.

 \begin{theorem}\label{thm:mainprobGab}
   Let $X\in \F_{q^m}^{k\times (n-k)}$ be randomly chosen. Then
  $$\Pr\big( \;\rs[I_k\,|\,X] \mbox{ is a gen.\ Gabidulin code }\big)\leq 
  \phi(m)q^{-(m-1)(n-k-1)(k-1)},$$ where $\phi$ denotes the
  Euler-$\phi$ function.
\end{theorem}

  \begin{proof}
    We have already seen in Lemma \ref{lem:probGabunion} that
   $$\Pr\big( \;\rs[I_k\,|\,X] \mbox{ is a gen Gabidulin code }\big)\leq 
   \mathop{\sum_{0<s<m}}_{(s,m)=1}\frac{|\mathcal
     G(s)|}{q^{mk(n-k)}}.$$ By Lemma \ref{lem:phi} part $3$, the sets
   $\mathcal G(s)$ all have cardinality $q^{k(n-k)}|\mathcal R_1^*|$,
   thus
   $$\mathop{\sum_{0<s<m}}_{(s,m)=1}\frac{|\mathcal G(s)|}{q^{mk(n-k)}}=\phi(m)\frac{q^{k(n-k)}|\mathcal R_1^*\cap \mathcal K|}{q^{mk(n-k)}}.$$
   Moreover by Lemma \ref{lem:R1}, we know that $|\mathcal R_1^*\cap
   \mathcal K|\leq (q^{m-1}-1)^{n-1}\leq q^{(m-1)(n-1)} $.  Combining
   all the inequalities implies the statement.
 \end{proof}

 % Observe that it is given by
%  $$ P[\mathcal G(1)]=\frac{|\mathcal G(1)|}{q^{mk(n-k)}}$$
%  so we are going to find a bound on
 
 We can now give the final main result of this work, that proves the
 existence of linear MRD codes that are not generalized Gabidulin
 codes for almost every set of parameters.
 
 \begin{theorem}\label{thm:main}
   \begin{itemize}
   \item For any prime power $q$, and for any $1<k<n-1$, there exists
     an integer $M(q,k,n)$ such that, for every $m\geq M(q,k,n)$,
     there exists a $k$-dimensional linear MRD code in $\F_{q^m}^{n}$
     that is not a generalized Gabidulin code.
   \item An integer $M(q,k,n)$ with this property can be found as the
     minimum integer solution of the inequality
     \begin{equation}\label{eq:main}
       1-\sum_{r=0}^kr\binom{k}{k-r}_q\binom{n-k}{r}_qq^{r^2}q^{-m}>(m-1)q^{-(m-1)(n-k-1)(k-1)}  
     \end{equation}
     taken over all $m\in \mathbb N$.
   \end{itemize}
 \end{theorem}

 \begin{proof}
   For fixed $q$, $k$ and $n$ consider the function
   \begin{align*}
     F(m) & =\sum_{r=0}^kr\binom{k}{k-r}_q\binom{n-k}{r}_qq^{r^2}q^{-m}+(m-1)q^{-(m-1)(n-k-1)(k-1)} \\
     & =aq^{-m}+(m-1)q^{-c(m-1)},
   \end{align*}
   where
  $$ a:=\sum_{r=0}^kr\binom{k}{k-r}_q\binom{n-k}{r}_qq^{r^2},  \;\;\;\;c:=(n-k-1)(k-1). $$ 
  % Note that $F(m)$ is the sum of the bounds from Theorems
  % \ref{thm:probMRD} and \ref{thm:mainprobGab}.
  Since $k\neq 1,n-1$, we have $c>0$. In this case $F(m)$ is the sum
  of two non-increasing functions and hence it is
  non-increasing. Therefore the function $1-F(m)$ is
  non-decreasing. Moreover it is easy to see that
  $$ \lim_{m\rightarrow +\infty} 1-F(m)=1.$$
  This means that the set of the solutions of Inequality
  (\ref{eq:main}) is non-empty. Then it has a minimum solution
  $M(q,k,n)$. Since the function $1-F(m)$ is non-decreasing, every
  $m\geq M(q,k,n)$ is also a solution of (\ref{eq:main}).  Hence, by
  Theorems \ref{thm:probMRD} and \ref{thm:mainprobGab}, we have the
  following chain of inequalities for every $m\geq M(q,k,n)$,
  $$\Pr\big( \rs[I_k\,|\,X] \mbox{ is MRD}\big)\geq 1-aq^{-m}>(m-1)q^{-c(m-1)}\geq \Pr\big( \rs[I_k\,|\,X] \mbox{ is gen.\ Gabidulin}\big), $$
  which concludes the proof.
\end{proof}

\begin{figure}[!ht]
  \begin{center}
    \includegraphics[width=7.25cm]{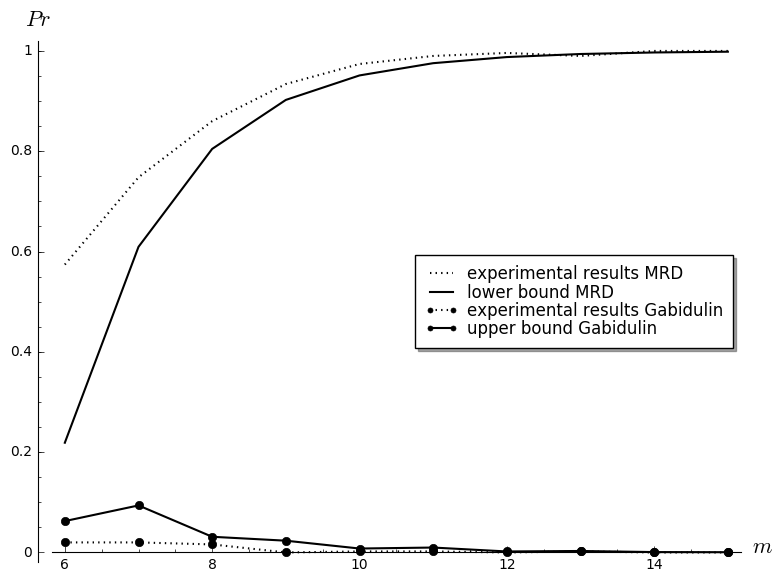}
    % \end{center}
    % \caption{Bounds and experimental results for MRD and generalized
    % Gabidulin codes in $\F_{2^m}^{2\times 4}$.}
    % \label{experimental1}
    % \end{figure}
%
    % \begin{figure}[!h]
    %   \begin{center}
    %     \includegraphics[width=7.25cm]{experimental-q2n5k2.png}
    %   \end{center}
    %   \caption{Bounds and experimental results for MRD and
    %   generalized Gabidulin codes in $\F_{2^m}^{2\times 4}$ and
    %   $\F_{2^m}^{2\times 5}$.}
    %   \label{experimental2}
    % \end{figure}
%
    % \begin{figure}[!h]
    %   \begin{center}
    \includegraphics[width=7.25cm]{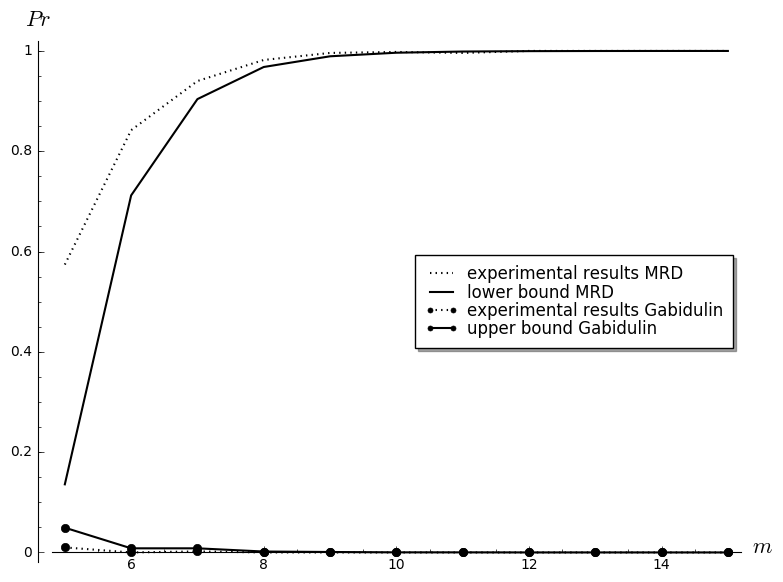}
  \end{center}
  \caption{Bounds and experimental results for MRD and generalized
    Gabidulin codes in $\F_{2^m}^{2\times 4}$ and $\F_{3^m}^{2\times
      4}$.}
  \label{experimental3}
\end{figure}

\begin{figure}[!ht]
  \begin{center}
    \includegraphics[width=7.25cm]{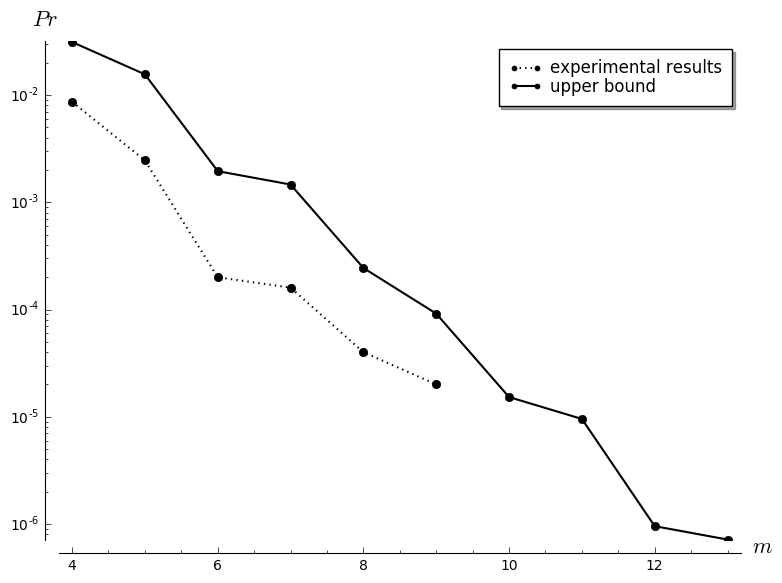}
    % \end{center}
    % \caption{Bounds and experimental results for MRD and generalized
    % Gabidulin codes in $\F_{2^m}^{2\times 5}$.}
    % \label{gabidulin1}
    % \end{figure}
%
    % \begin{figure}[!h]
    %   \begin{center}
    \includegraphics[width=7.25cm]{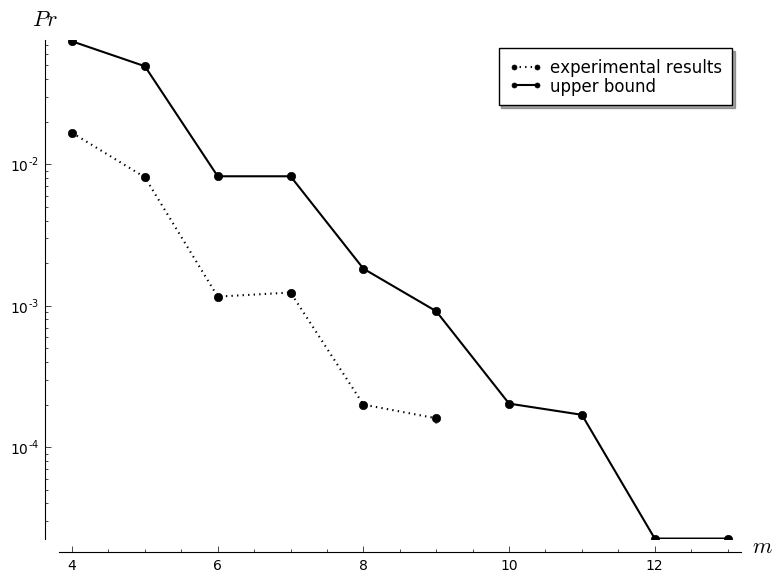}
  \end{center}
  \caption{Bounds and experimental results for generalized Gabidulin
    codes in $\F_{2^m}^{2\times 5}$ and $\F_{3^m}^{2\times 4}$.}
  \label{gabidulin2}
\end{figure}

In Figures \ref{experimental3} and Figures \ref{gabidulin2} we compare
the bounds derived in this section with experimental results, which we
got by randomly generating over $500$ rank-metric codes. The
continuous lines show the bounds, the dotted lines show the
experimental probabilities. In Figure \ref{experimental3} we see that
Gabidulin codes are very few among all MRD codes when the extension
degree $m$ is large.  The probabilities for generalized Gabidulin
codes decrease so quickly for increasing parameters that we show them
separately, in logarithmic scale, in Figure \ref{gabidulin2}. Notice that from $m=10$ it is very difficult to generate a generalized Gabidulin code randomly and thus, experimentally we got a probability zero. This is why the experimental result was shown only up to $m=9$.

% To see that Gabidulin codes are very few among all the MRD codes
% when the degree of the extension is large, we can look at the
% Figures \ref{experimental1}, \ref{experimental2} and
% \ref{experimental3}. The continuous lines show the bounds for the
% probability of MRD codes and the probability of generalized
% Gabidulin codes. The dotted lines show the experimental
% probabilities which we got by randomly generating over 500
% rank-metric codes. These figures also show how far our bound is from
% the real values of the probabilities. The probabilities for
% Gabidulin codes is figures \ref{experimental2} and
% \ref{experimental3} are not visible enough so we also show them
% separately in figures \ref{gabidulin1} and \ref{gabidulin2} but in
% logarithmic scale for the probability. Notice the oddities in the
% graphs, which are due to the fact that when $m$ is large, it is hard
% to generate a generalized Gabidulin code randomly.

%%%%%%%%%%%%%%%%%%%%%%%%%%%%%%%%%%%%%%%%%%%%%%%%%%%%%%%%%%%%%%%%%%%%%%%%%%%

\section{Conclusion}\label{sec:conclusion}

In this work we have shown that, over the algebraic closure of a given
finite field, MRD codes and non-Gabidulin codes are generic sets among
all linear rank-metric codes. For this we have used two known criteria
for these two properties, which give rise to algebraic descriptions of
the respective sets. Afterwards we have used the same two criteria to
establish a lower bound on the probability that a randomly chosen
systematic generator matrix generates an MRD code, and an upper bound
on the probability that a randomly chosen systematic generator matrix
generates a generalized Gabidulin code. With these two bounds we were
then able to show that non-Gabidulin MRD codes exists for any length
$n$ and dimension $1<k<n-1$, as long as the underlying field size is
large enough.

\bibliography{./network_coding_stuff}
\bibliographystyle{plain}

\end{document}